\documentclass{vgtc}                          % final (conference style)
\ifpdf%                                % if we use pdflatex
  \pdfoutput=1\relax                   % create PDFs from pdfLaTeX
  \usepackage{graphicx}                % allow us to embed graphics files
  \DeclareGraphicsExtensions{.pdf,.png,.jpg,.jpeg} % for pdflatex we expect .pdf, .png, or .jpg files
\else%                                 % else we use pure latex
  \usepackage{graphicx}                % allow us to embed graphics files
  \DeclareGraphicsExtensions{.eps}     % for pure latex we expect eps files
\fi%

\graphicspath{{figures/}{pictures/}{images/}{./}} % where to search for the images

\usepackage{microtype}                 % use micro-typography (slightly more compact, better to read)
\PassOptionsToPackage{warn}{textcomp}  % to address font issues with \textrightarrow
\usepackage{textcomp}                  % use better special symbols
\usepackage{mathptmx}                  % use matching math font
\usepackage{times}                     % we use Times as the main font
\usepackage{cite}                      % needed to automatically sort the references
\usepackage{tabu}                      % only used for the table example
\usepackage{booktabs}                  % only used for the table example
\onlineid{0}

\vgtccategory{Research}

\vgtcinsertpkg

\title{What Do We Actually Learn from Evaluations \\ in the ``Heroic Era'' of Visualization? \\[1.25em]
\Large{Position Paper}
}

\author{Michael Correll\thanks{e-mail: mcorrell@tableau.com}\\ %
        \scriptsize Tableau Research %
}

\abstract{
We often point to the relative increase in the amount and sophistication of evaluations of visualization systems versus the earliest days of the field as evidence that we are maturing as a field. I am not so convinced. In particular, I feel that evaluations of visualizations, as they are ordinarily performed in the field or asked for by reviewers, fail to tell us very much that is useful or transferable about visualization systems, regardless of the statistical rigor or ecological validity of the evaluation. Through a series of thought experiments, I show how our current conceptions of visualization evaluations can be incomplete, capricious, or useless for the goal of furthering the field, more in line with the ``heroic age'' of medical science than the rigorous evidence-based field we might aspire to be. I conclude by suggesting that our models for designing evaluations, and our priorities as a field, should be revisited.
} % end of abstract

\CCScatlist{
  \CCScatTwelve{Human-centered computing}{Visu\-al\-iza\-tion}{Visualization design and evaluation methods}{}
}

\begin{document}


\firstsection{Introduction}

\maketitle

%% \section{Introduction} %for journal use above \firstsection{..} instead
The late 18th century began what is often referred to as the ``heroic age'' of medicine in the west~\cite{sullivan1994sanguine}. Doctors would wade into battlefields or epidemics, roll up their sleeves, and prescribe mind-boggling amounts of diuretics and purgatives and mercury, all the while bleeding patients repeatedly. It was not particularly effective medicine, but one of its effects was to move medical practice firmly into the realm of the ``expert.'' Before, anybody could read their Pliny the Elder, see that the cure for typhus was (say) a poultice of goose grease and honey, and apply it (the patient would still probably die, but, you know). Now, even for minor illnesses, I needed to consult with an expert to see exactly how much bloodletting needed to be done, and at what tempo.

We're firmly in the heroic age of information visualization. We don't have much of a theory, and what we have might be grossly incorrect, but, by gum, we're gonna get things done. Take two bar charts and call me in the morning. What this has created is an odd incentive where lots of us are heading out to work with ``domain experts,'' make a new visualization system, do some sort of evaluation of the system to confirm that it works, and then write it up. Did they really \textit{need} a new system? If they did, was it really worth spending several months of back and forth meetings and prototyping and design work? These are the wrong questions for the heroic age of data. Our lack of strong theoretical unpinnings, combined with our emphasis on \emph{techne} (``knowing-how'') rather than \emph{episteme} (``knowing-that'')~\cite{parry2014techne} means that our incentives for evaluating visualization systems are often quite perverse and fail to tell the broader community much of interest. Most troubling, like the blood-letting and purgatives and diuretics of old, our lack of solid  epistemological foundation mean that we can \textit{remain in error} about the efficacy of our work for long periods of time.

In this setting, I believe that our current practices around evaluations do not suffice to reliably assess the contribution of any particular work to the field. I will attempt to convince the reader of this ontological insufficiency through a series of thought experiments. Note that is largely a claim about logical necessity and sufficiency, and so these thought experiments may or may not have much to do with empirical lived reality. I acknowledge that determining \textit{how often} we fall into the particular pitfalls I will bring up is an empirical project I am unwilling to undertake for this paper. Also note that this argument is mostly about evaluations conducted as part of design studies or other evaluations of visualization systems. Graphical perception work, sensitivity analyses, and other parts of empiricism are out of scope (although I do think that some of the issues I raise could apply to our practices around these sorts of studies as well).

\subsection{A Sketch of the Argument}
The procedural \textit{techne} focus of knowledge-formation (as well as our generally \textit{positivist} epistemology~\cite{meyer2019criteria}) in visualization design extends also to how we design our evaluations. Tamara Munzner~\cite{munzner2008process} suggests an injective matching procedure from the \textit{type} of contribution we intend to make and the resulting appropriate forms of evaluation and evidence. Lam et al.~\cite{lam2011empirical} similarly divide visualization evaluations into seven core \textit{types} with accepted forms, measures, and analyses. These procedural approaches to visualization evaluation suggest to me the following tacit premises:

\begin{enumerate}
    \item The \textit{kind of work} we will do to build our visualization system suggests the \textit{kind of evaluation} we ought to undertake, as well as which \textit{metrics} we should collect, \textit{a priori}.
    \item This evaluation can \textit{succeed or fail} in illustrating the utility of our system by our chosen metrics.
    \item The \textit{success or failure} of the evaluation is \textit{diagnostic or informative} to the contribution of our work to the field.
\end{enumerate}

These premises are at some level vacuously true. For instance, we do not need to finish the implementation of a newly developed algorithm to know that, say, a Tarot card reading would be a poor fit for evaluating the algorithm's average run time compared to a quantitative performance evaluation (except under very specific circumstances~\cite{mcnutt2020divining}), and that this quantitative evaluation could produce either good or bad news for the algorithm compared to the current state of the art, and that a paper with this good or bad news would help me judge the contribution of the algorithm to the literature in a more precise way than a paper with no such information. Nevertheless, I think we adhere too strongly to these premises when considering our evaluative work in the contexts of design studies. In particular, I maintain:

\begin{enumerate}
    \item A visualization system can be ``good'' (in that it can succeed by the reasonable metrics we laid out in our evaluation) and still be largely uninteresting to the field.
    \item Conversely, a visualization system can be ``bad'' (from the same evaluative standpoint) and be very interesting.
    \item Therefore, our evaluations (even ones that are a good ``fit'' for our intended contribution) may only tell us whether something is ``good'' or ``bad'' rather than ``interesting'' or ``uninteresting'' (and even then only in a narrow and stochastic way).
\end{enumerate}

It follows from the above that \textit{the outcomes of evaluations, even appropriately-designed evaluations, may be uninformative for assessing visualization systems from a research perspective}. Evaluations, even well-designed evaluations that follow established norms and rules for the sort of systems work we did, are not magical procedures that lend the imprimatur of seriousness or utility to our visualization papers, and we as researchers or reviewers should give them weight only insofar as answer our questions about the work in a rhetorically convincing way. 

\section{The Multiple Worlds of Visualization Evaluation}
In order to disentangle the \textit{fit} and \textit{outcome} of an evaluation from the \textit{research contribution} of the visualization artifact being evaluated, I will rely on a form of thought experiment that Daniel Dennett calls an ``intuition pump''~\cite{dennett2013intuition}. In this case what I will be doing is taking a simple premise about which we have a similar set of intuitions and adding increasing layers of complexity or absurdity to see where our intuitions begin to shift. Again, this is a thought experiment: the practical \textit{likelihood} of any particular outcome here is sort of a red herring (much as it is sort of besides the point to suggest if Schrödinger's cat is a calico or a tortoiseshell). What I hope to do instead with these scenarios is to cast doubt about many of our existing beliefs about user evaluations, e.g. that ``good'' outcomes vis-à-vis metrics like user performance, user satisfaction, or analytical insight, are indicative of a visualization system of research interest.

That prelude out of the way, let's begin with the following prompt, right out of the heroic age of visualization:

\emph{You are a visualization researcher running a research lab. Your lab regularly submits to top-tier conferences like IEEE VIS or CHI or what have you. You've just been offered the opportunity to work with a group of cancer researchers who have been struggling to understand their data. After a long period of collaboration and iterative design sessions, you come up with a visualization system, \textit{CancerVis}. Using your system, the researchers discover something that leads them to develop a cure for cancer. There's a press junket, a Nobel Prize or two, and the scientists make sure to thank you in their acceptance speeches.}

\emph{Meanwhile, you've had a student or two working at evaluating your system via your favorite evaluative method (quantitative, qualitative, insight-based, whatever floats your boat). Today, you've just received an email with the ominous subject heading ``RE: Study Results.''}

Let's pause here for a moment. Right now, given only the information in the story I've just told you, think about the truth or falsity of the following statements (subjective though they might be):

\begin{enumerate}
    \item CancerVis is a good visualization system.
    \item A paper about CancerVis deserves to be accepted at a high tier visualization conference or journal.
    \item The visualization community can learn from CancerVis.
\end{enumerate}

Jarke van Wijk, in his consequentialist assessment of visualization~\cite{van2005value}, would seem to produce the inescapable conclusion that proposition 1 is true: millions of lives saved is certainly worth whatever cost in design/training/adoption of a team of a few scientists and designers. Unless you're extremely uncharitable with my story above (and we will be later, don't worry), it's hard to argue for anything other than our system's positive utility.

Propositions 2 and 3 seem dicier to me. I've told you nothing about the actual CancerVis system. I have no idea what, if anything, was new about it. It's possible that any half-competent designer would have produced a system just as good (or even better) than CancerVis. It's possible that your methods and designs were so shoddy that they should only be brought up in academic circles as a warning about what not to do. I just don't know. It's likewise hard to say if a paper based on CancerVis deserves to be accepted in a top-tier venue, just in the same way that doctors don't get a \textit{Lancet} article every time they treat a patient who doesn't die. I don't know, \textit{a priori}, whether CancerVis has any lessons that would advance the field, which is allegedly one of the main things that academic papers are supposed to do.

I would like the clarify here that my question about ``deserves'' is on the wrong side of an is/ought distinction. I know that I would feel like a pretty big jerk if I didn't let the people who helped cure cancer at least come to the conference and give a talk. So I'm pretty sure that a CancerVis paper would, in practice, almost certainly be accepted if it were pitched the right way (it's a hell of a ``broader impacts'' statement, at least). But right now, without further information, I don't know if the process of making and deploying CancerVis generated generalizable knowledge for the field of visualization.

Given this current state of affairs, let's open that ``RE: Study Results'' email. We'll be exploring different parallel worlds, some of which are mutually exclusive (and some admittedly fantastical). After each possible world, I would charge the reader to reflect on how their initial conceptions about the system and its benefit to the field have (or have not) changed. Here's the first such world:

\subsection{The Unique System}

\emph{The students running your evaluation found that CancerVis beats all of the other systems in this space totally out of the water. That was easy enough to show, since there were no other systems in this space. The problem that it solved is idiosyncratic to your specific set of domain collaborators. ``And we're out of the game,'' said the lead scientist. ``We cured cancer so we're all going to retire to tropical islands and live it up. Have fun with your scatterplots or whatever,'' they continue, packing their belongings into a cardboard box while checking out yacht prices online. Those oncology research labs that remain have dramatically different data problems that your system can't address.}

By construction, it seems like CancerVis is a good system. But it is also, by construction, unclear if the VIS community would actually learn anything from it. It solves one problem, and did so with such definitiveness that nobody has to solve the problem again. And, to extend the medical metaphor a bit more, it's a ``zebra'' problem, not a ``horse'' problem~\cite{sotos1991zebra}. So why would anybody in the field consult a paper about CancerVis paper? What use would they get out of it? CancerVis is nice, and the designers deserve all sorts of accolades in this particular version of the world, but I'm not certain an academic paper is the right carrot for this work. Or perhaps I'm being too negative, and the fact that you followed a design procedure that led to a good outcome is itself good to know. Let's take that situation to the extreme in the next world we visit.

\subsection{The Obvious System}
The prior situation was perhaps a little unfair. Given the level of abstraction involved in visual analytics task analysis, it's unlikely that the problem CancerVis solved was just totally out of left field and has nothing that could be transferred to a different problem. So let's move to a different world:

\emph{The students running your evaluation found that the system works just fine. But, then again, it ought to work fine. Your students read a popular visualization textbook and then followed the procedures in it to the letter. ``We had categorical data with associated aggregate quantitative data, so I used a bar chart for that bit,'' says the main designer. ``We had time series data so we used a line chart,'' continues another. As a test, you give the task requirements and sample data to your undergraduate visualization class as a design exercise. They all came up with designs almost identical to CancerVis, except for one group in the back who went with a 3D pie chart where each slice is a word cloud, but you think they might have been messing with you.}

Now, we've sort of got the opposite problem from the Unique System. Here, our problem was a little too easy, such that almost everybody with a perfunctory understanding of visualization design could tackle it. It's good to know that what we think works actually works (especially since our basis for believing such things is often rather thin~\cite{kosara2016empire}), so the generalizable knowledge we get from this would seem to be, largely, to keep on trucking.

If so, there would then seem to be diminishing academic returns for such papers, except in the aggregate, as a way to perform meta-analysis and see what accepted design practice looks like. We might remember Alexander Fleming, for instance, for discovering penicillin, but we likely don't remember the hundredth or thousandth person to prescribe penicillin. We would only seem to need to know about it in a man-bites-dog situation: when the things we expect to work from our standard design processes stop working (say, whatever the metaphorical equivalent of drug resistance is). Let's visit one such world next:

\subsection{The Worse Than Baseline System}

\emph{The students running your evaluation have found that the main insight that led to the breakthrough discovery would have shown up in an Excel pivot table. In fact, in their evaluation of the system, they found that people were much faster in finding it with Excel than with the bespoke CancerVis system. Your collaborators were just about to try tinkering with Excel more, but your initial planning meeting lo those many months ago interrupted them, and they figured you would know what you were doing. Extrapolating from that performance data, it's possible that the intervention of your lab delayed the cure for cancer by months.}

Yikes! Now we've reversed the sign of that utility equation in Van Wijk's Value of Visualization assessment I mentioned to above. We showed up, did (presumably) rational things, and it led to a worse outcome than if we hadn't intervened at all. We didn't help people. CancerVis is a bad system. But here a post-mortem might be useful. Why was it bad? Was it too complicated? Poorly designed? If the message of the CancerVis paper is ``I personally am incompetent and didn't know what I'm doing'' then maybe that's less valuable for the field, but if the message is ``we did all the right things and still got this bad outcome'' then we've got something here.

I should note, however, that I had to set up this world \textit{very} carefully just so this assessment of failure was even \textit{possible}. There's a baseline system (Excel) to compare against, an initial plan of attack that was derailed, and a task resulting in an insight that was modeled in sufficient detail to use as a yardstick. In most visualization system evaluations, we have no idea whether or not we're in this world, or one like it. Our collaborators' problems are idiosyncratic enough that we don't have a baseline to test against, and what is meant by ``success'' is nebulous enough that it's hard to pin down. It should perhaps worry you that we almost never test against visualization ``placebos'' in this way. We often have no way of knowing if what we're doing is any more or less helpful than an 18th century doctor proscribing daily purgatives.

\subsection{The Detestable System}

\emph{The students running your evaluation found that your system works fine, but your users hated using it the entire time. Un-intuitive interfaces, hostile design choices, and a color scheme that led to migraines in some of your users after just a few minutes. Your student collected some qualitative feedback, and a lot of it is just direct threats of violence against you personally. That being said, they did perform better with your system compared to the baselines.}

This particular world might be the most far-fetched of all. Due to the demand characteristics~\cite{orne1962social} of how we run our experiments (where we often form close friendships or working relationships with our collaborators and have mutual stake in the others' successes) as well as the good old sunk cost fallacy, there is a pressure to please the experimenter and give positive feedback even if the system is awful~\cite{dell2012yours}, or to persevere with less than ideal systems to keep a collaborative relationship alive. As with the prior world, I would ask you to consider how many things would have to be true (outspoken collaborators operating in an environment of radical candor, mixed methods quantitative and qualitative methods against existing baselines, etc.) in order for you to ever find out that you were in a world where people really didn't like the thing that you built.
But suppose we were in such a world, where it's clear that, from a human-centered design standpoint, we did almost everything wrong. And yet, the users of our tool did cure cancer. And the quantitative feedback does seem to show some performance benefits for our system. Given all of that, do you care that your users hated it? The unfriendly nature of your design might make it an uphill battle to productize or sell the system, and perhaps tricky to maintain a working relationship with the oncology team after the CancerVis project is over, but for the purpose of ``getting a paper or two'' out of the thing, you could argue that this negative qualitative data simply doesn't matter.

This seems like a counter-intuitive conclusion to me. The Kantian in me would say that performing human-centered design work is an end in and of itself. And I would expect that, as a general rule, a human-friendly system would be ``better'' by most quantitative analytical metrics (engagement, willingness to explore, etc.) than a functionally equivalent system that is human-hostile. But what should I do with these particular results, other than writing yet another paper about the existence of a preference/performance gap?

\subsection{The Serendipitous System}
Now let's jump away from that sad realm to a different parallel world, and open that email up again:

\emph{The students running your evaluation found that your system was pretty good at the tasks that were initially given to you, but that those tasks had almost nothing to do with how the breakthrough came about. The main insight was found almost totally by chance. Your collaborators performed precisely the right sequence of actions in precisely the right order. If they had set some sliders differently, say, there's a chance they would not have found the key insight at all.}

There is a concept in philosophy called ``moral luck,''~\cite{williams1976moral} where our intuitions about praise and blame in moral actions seem to be partially reliant on chance. For instance, the criminal penalties for attempted murder are often lower than for ``actual'' murder, even if the actions and intents of the person doing the violence were the same. Here we seemed to have lucked out.

It's not clear, however, what we've learned from this system, other than ``sometimes you luck out.'' Now, there are ways of designing to promote this kind of serendipity~\cite{alexander2014serendip,thudt2012bohemian}, but that doesn't seem to be what happened here. And what that means is that if we were evaluating a system based on its overall benefit, we'd get radically different answers depending on the luck of our participants.

There's a sort of related issue here in that nothing succeeds like success (or, more negatively, the rich get richer). The people who are most likely to generate their own analytical luck are highly motivated people who are willing to give you the benefit of the doubt and spend lots of time with you and your project. Those people would self-select for visualization collaborators with a proven track record of success. So successful visualization labs are perhaps more able to self-select (or ``winnow'' down~\cite{sedlmair2012design}) successful domain collaborators who are more likely to get something of value out of whatever system they are given, no matter the quality, which means that the visualization lab gets even more successful, and so on and so on. How important is first mover advantage in a field like visualization?

But matters of inequality of accolade distribution aside, at the very least I would hope that this version of the world would indicate that an anecdotal incident of a successful use of a particular visualization system may not be a strong case for its general utility. I'm sure at least some of the people given mercury and blood letting did get better (if for no other reason than their bodies were able to fight off the disease by themselves despite the ``help'' of their physician). But one healthy patient does not mean we know what we're doing. One happy analyst does not mean we've built a good system. Maybe we were just lucky.

\subsection{The Super Serendipitous System}
A natural objection to the world I just presented is that the designers built a system that ``let accidents happen,'' and afforded the kind of exploration that made the insight possible. That, you might think, is at the very least evidence that they were doing something right. So let's move on to another, highly related parallel world:

\emph{The students running your evaluation have found that the main insight was found due to series of errors in an earlier version of your system. It made a bar that was supposed to be \emph{blue} render as \emph{red}, which the collaborators found odd enough to investigate. Then the system crashed to desktop, which gave the scientists some time to think about what would happen if that bar actually \emph{were} red. This led them down a path that unravelled the whole problem. If you had done a better job at software engineering, then they might have never found anything worth talking about.}

Now moral luck is really at play here. Your group did a bad job that just so happened to work out. Your system was sort of the equivalent of the (somewhat apocryphal) story of Fleming forgetting to clean up the bread mold in his lab. I'm sure these sorts of errors happen all the time, especially over the course of iterative development, but it's not clear what all I'm supposed to learn from any of this. ``Hey, sometimes things just work out, despite our actions'' is not really a contribution statement I can do much with. CancerVis, in this world, would be an interesting anecdote, but I'm not sure it would make a good model for anything.

\section{Discussion}
To me, these multiple worlds produce an inescapable conclusion. If I'm an academic, and I'm supposed to be assessing a paper based on its academic contribution to the field: \textbf{
Whether your system ``works,'' in the sense of being well-designed or useful or well-received, might be the least interesting thing about it.
}

It's nice if it worked to solve a problem. And it's admirable if you took the time to confirm that people liked it. And I deeply suspect that soliciting iterative feedback and testing different designs and features will help you do a better job at engineering the dang thing. But from an academic perspective of contributing to the field, there's a good chance that I don't (or shouldn't) care. I am mainly interested in the answers to these questions (or ones very similar to them):
\begin{enumerate}
    \item What does your system tell me about \textbf{visualization design}?
    \item What does your system tell me about \textbf{people}?
    \item What does your system tell me about what we should \textbf{do next}?
\end{enumerate}

In short, what do I know that I didn't before, and what should I do now that I know it? In the worlds above there are potentially answers to all of those questions. But they are very different answers depending on the world we are in, and the type of analysis we did. The existence of an evaluation \emph{per se} does not help me narrow things down, or automatically strengthen the contribution of the paper in a rigorous way. Nor is it necessarily a question of the ``fit'' of evaluative methods to kind of work we did (after all, we did roughly the same work in every single parallel world). It seems to be more about the fit of the evaluative methods and the kinds of things we want to say.

In other words, the ``standard'' design study procedure of finding a domain collaborator, building a tool to solve their problem, evaluating the tool, and then writing a paper about it, doesn't necessarily advance the field, even if the domain experts had interesting problems and there is some empirical evidence that the thing we built was beneficial via some metric. It might be useful for us \textit{personally} (or even \textit{organizationally}) to build experience engineering software, discover hitherto unknown design pitfalls, and meet new people with new kinds of data, but this design study pipeline might not move the field forward one iota.

\subsection{Counter-Arguments \& Case Studies}
The perfect is famously an enemy of the good. An uncharitable reading of my thought experiments would suggest that not only must an evaluation be well-designed for me to take it seriously, but that it must be well-designed across an arbitrary number of possible worlds, some of which are quite unlikely and adversarial. While it's true that I wouldn't say no to analyses that can survive across the multiverse (as in Dragicevic et al.~\cite{dragicevic2019increasing}), my thought experiment was more designed to illustrate ways in which a properly performed evaluation can still fail to teach us something useful as a field. Natural counterpoints to this argument are that 1) in an inductive sense, just because we didn't \textit{necessarily} learn something useful doesn't mean that we didn't \textit{potentially} learn something useful (who could predict which papers at which conferences will win something like an IEEE VIS Test of Time award, after all) and 2) we may not have learned anything useful from one study \textit{per se}, but we might learn something from studies in the \textit{aggregate}. 

I think these objections have a similar form: that we should be accepting of large numbers of papers with the expected value of each paper in terms of generating new knowledge for the field being potentially quite small. In none of the parallel words I bring up, after all, was the evaluation \textit{completely} uninformative. We uncovered \textit{some} new information from each of them, even if the information wasn't particularly actionable (``we screwed up and it worked out anyway'') or novel (``if you follow a similar design process to everybody else, you'll get similar results'').

The question would then seem to hinge on whether we learned \textit{enough} in each world to ``deserve'' a paper (moving through the stages of grief for our evaluations from denial to anger to bargaining in record time). To correct my earlier statement, I will mention that a doctor probably learns \textit{something} every time they treat a patient, but they don't always learn enough that they go shouting from the rooftops about it (although I should point out that most hospitals \textit{do} hold morbidity and mortality conferences when patients \textit{die}). Doctors \textit{do}, however, write lots of case studies. I think it is in the analogy to case studies where we can explore this counter-argument to the fullest.

The history of psychology and medicine is incomplete without a gallery of influential patients. Stories of Phineas Gage~\cite{damasio1994return} and Genie the ``feral child''~\cite{curtiss1974linguistic} are told in undergraduate courses to illustrate foundational points about how the mind works. Patients like Alexis St. Martin (whose recovery after a wound left a permanent fistula that could be used to study in great detail the workings of the digestive system~\cite{bensley1959alexis}) and Sadao Yoshida, who voluntarily consumed parasitic roundworm eggs in order to confirm hypotheses about their life-cycle~\cite{yoshida1919development}, were able further the cause of common scientific knowledge through self-sacrifice. Those extremes aside, even today, case studies on single patients are ubiquitous in medicine.

At a glance, a typical design study is a lot more like a case study. The evaluation may say $n=500$ Mechanical Turk workers or what have you, but it's still fundamentally an $n=1$ study with the ``patient'' being our collaborators as a gestalt and our ``treatments'' being the design(s) we gave them. The evaluations we perform at the end are perhaps not meant to answer generalizable questions about the field, but are really just diagnostic tools like a stethoscope or a thermometer: we're just ``checking the vitals'' on our patient to see if our treatment helped them our not.

In the case study regime, the research contribution of any individual paper might be rather small. They are ``anecdata'' that are meant to be either a) existence proofs (``The patient presented with symptoms that were unlike any I have seen, which is evidence of a new disease'') b) hypothesis creation steps (``I recommended this course of treatment for the patient and it seemed to work, we should study this in more detail to confirm'') or c) the raw material for later meta-analyses (``I looked at a hundred patients that were given a new treatment and they seemed to have better outcomes than others''). 

If we are in this regime, then my suggestions that design study evaluations don't provide us with much in the way of generalizable knowledge is perfectly fine (or at least generally okay). Individual design studies papers are not meant to be of much interest, but packs of them together, analyzed in aggregate, will tell us new things. We have as a field begun to undertake these sort of systematic reviews. We have collections and browsers of the dozens to hundreds of techniques or systems for visualizing trees~\cite{schulz2011treevis}, texts~\cite{kucher2014text}, time~\cite{tominski2017timeviz}, and uncertainty~\cite{jena2020uncertainty}. Perhaps more to the point, we have also begun to collate evaluations of designs, either in general~\cite{lam2011empirical}, or in specific cases of glyphs~\cite{fuchs2016systematic} and uncertainty visualization~\cite{hullman2018pursuit}.

I have two responses to this point of view, one flippant and one more involved. The more flippant one is, if a design study is just meant to be a point example, \textit{why bother evaluating at all}? If you're giving up on using the evaluation to generate generalizable research knowledge, but just (in a metaphorical sense), to confirm that the patient recovered (or at least didn't die), why spend so much time and effort on evaluating, or critiquing evaluations, or demanding them as reviewers? Early influential papers (like those for treemaps, say~\cite{johnson1999tree}) didn't have any user evaluations, so why do we think we need them now? As existence proofs of new domains or new potential designs all we need is just some evidence that they do \textit{something}, and then we can rely on future work to figure out if that something is useful, or consistently useful, or better than some other design.

My other response to the medical case study metaphor is that \textit{if we're meant to be using case studies as anecdotes for later meta-analysis, we sure don't write them in a way that would be useful for that purpose.} Quantitative studies still only occur in a minority of VIS papers~\cite{harozperformance2017}, we only infrequently share our results in an open and accessible way~\cite{haroz2018open}, and, if my experience with uncertainty visualization meta-analysis~\cite{hullman2018pursuit} is any indication, we are wildly inconsistent with what we measure and how we measure it in the first place. These idiosyncrasies mean that I'm afraid we can't just paper over deficiencies in how we evaluate now by appealing to some potential research contribution in the future: at the very least it seems rude to future generations of researchers to make them have to pick through the rubble of our current practices to find the few apples-to-apples comparisons they can salvage.

In summary, there's nothing wrong with writing a case study. They are often interesting to read, teach us about a new domain we might not have heard of (I suppose these are the ``new disease spotted'' equivalents), and are breeding grounds for new designs. I would however ask for two reforms: 1) that we be clear-eyed and honest when we design our evaluations and write them up: we aren't setting out to confirm universal truths about human reactions to visualizations, we're just showing that our design seems to do what we claimed it does, which may not require any sort of quantitative evaluation at all and 2) that we make our papers ``talk to each other'' better: use standard metrics when we can, avoid the idiosyncratic and often impenetrable ``task analyses'' that generate the $n=1$ paper experimental conditions for our work, and rely on open data practices to make meta-analyses possible, easier, and more useful.

\subsection{What is to be Done?}

Here I revisit my initial metaphor of visualization's ``heroic age.'' To me our current heroic era of visualization is characterized similarly to the heroic age of medicine:

\textit{An emphasis on individual herculean actions by individual actors.} Many of our design study papers focus on how difficult it was to get the right data in the right format, or to create the right designs, or to foster the right sort of collaborations. The assumption in such papers, tacit or explicit, is that other labs or ways of thinking would not have produced the same positive results. Our visualization ``heroes'' are often some combination of clinician, evangelist, and engineer: the first to intervene in specific domains hitherto unreached by academic visualization, the first to crack the puzzle of the ``right'' way to collaborate with this strangers, or the first to make the ``right'' kinds of tools. As our field matures, we will need other ways of evaluating our contributions other than these appeals to novelty and individual insight.
 
\textit{A lack of ``safe'' placebos and interventions.} Many of the medical interventions after the end of the heroic age were equally as (in-)effective as mercury and bloodletting, but, e.g., hydrotherapy and fad diets and other such placebo treatments were much less harsh on the system than weeks of purgatives, and so were preferred as interventions by patients and clinicians alike. 
A century or so after the start of heroic age, Flint speaks of a 19th century turn from ``heroic practice'' to ``conservative medicine''~\cite{flint1874essays}) that is cognizant of the size, scope, and potential disruption of the intervention, favoring the safer intervention when possible. By contrast, our usual intervention (the design and implementation of bespoke visualization tools through iterative and collaborative design) is tremendously expensive in terms of time, energy, and effort. Making an entirely new tool I view as a very radical act in many other scientific fields; in visualization and HCI it seems to be the norm.

\textit{A lack of theoretical correctives.} Doctors of the heroic age like Benjamin Rush did not lack theory. Their actions were often strongly situated within humorism, at that point a theory with over 2,000 years of application to medicine from Hippocrates on down. What was lacking, in my view, was a willingness to revisit this theory (rather than for instance claiming that a patient died after blood-letting because we didn't bleed them \textit{enough}), the epistemological tools to create new theory, and the empirical and rhetorical tools to supplant the old theory with the new (here I spare a prayer for Ignaz Semmelweis, who was unable to convince his colleagues of the importance of antiseptic handwashing, and died in an asylum after widespread mockery~\cite{best2004ignaz}). 

If we really are in the heroic age of visualization, and our focus is on repeated practice and intervention rather than theorizing and verification, then much of our systems work is not going to tell us much, with or without an evaluation. If we want to focus on evaluations in our work then we need to correct one or more of the potential problems of our heroism:

We write too many systems and design study papers and \textbf{we need scholastic and academic rewards for visualization design that are not conference papers}. Again, doctors don't get a top tier paper in a medical journal every time they treat a patient who doesn't die. This is a generalization of course, but we as an academic field seem to have somewhat similar design patterns and principles for building and evaluating visualization systems (or at least idealized forms of these practices embodied in influential papers such as Munzner et al. and Sedlmair et al.~\cite{munzner2009nested,sedlmair2012design}). If this is the case, then ``we followed the process and it worked out'' is important to know in a general way (just as it is important to know that the sun continued to rise in the east today), but it isn't really enough to hang our hats on (although if we see that the process \textit{isn't} working, that is useful information). Yet I acknowledge the practical utility of doing these design studies in academia: to build personal or institutional know-how, to provide deliverables to keep collaborators happy, to stake one's claim to expertise in a particular data domain, etc. 

I agree that we should reward practical design work, encourage grad students to build systems and hone their design skills, and record what systems we built and why (so we can perform meta-analyses or develop best practices), but there \textit{has to} be a reward structure for doing this other than building hundreds or thousands of independent visualizations systems, each with independent, largely un-comparable evaluations, all of which are published as 8-10 page conference papers. Before I feel comfortable suggesting altering incentives here (one unpalatable solution to this problem would be to just simply reject every design study paper that doesn't teach me anything useful for visualization), I think getting these alternative reward structures in order is important. For instance, we could make more of a habit in the field of publishing our design study work in the journals or venues of our domain collaborators (after all, if we claimed to have helped our users, then we should tell the rest of the domain that so they can be helped as well). Or we could encourage the building of personal or lab-wide ``portfolios'' as with art or design schools rather than more traditional paper-based CVs (and evaluate students and colleagues on this basis rather than just paper count).

We are too concerned with the ``success'' of our designs and lack adequate appreciation for failures, or concern with alternatives to our own efforts. \textbf{We need more equivalents of ``visualization placebos,'' and greater willingness to detect (and report on) our design failures}. We assume that, as visualization designers with expertise, all of our interventions will eventually be successful (maybe after enough design iterations). This may or may not be the case, but there's often no way to tell. We should be willing to cut our losses and report out on what didn't work. And a ``loss'' here may have little to do with how a system is received or functioned: as researchers we presumably seek out systems work with domain collaborators with the goal of solving mutual interesting research questions. It's possible that we completed the project having satisfied the domain scientists, but without having learned very much in the way of new, generalizable knowledge for our home discipline. Alternative venues to showcase our ``failures'' such as the Fail Fest workshop (\url{https://failfest.github.io/}) might provide some ways to showcase these missteps, but I think we should do more to actively seek out potential points of failure in our own research. This could involve adversarial analyses (the equivalent of visualization ``red teams,'' perhaps) post-paper publication post mortems (how many visualization creators of one-off systems for domain experts check in to see if their collaborators are still using them 1, 2 or 10 years down the line?), or re-analyses and replications of our results by other groups.

We've gotten too far ahead of ourselves, and \textbf{we need more theoretical underpinnings, meta-analyses, and codifications of standard practices before we do more practical work.} A quantitative evaluation could tell me which of the four humours was most effective in treating the flu, but it would be less useful in telling me that the whole regime of the humorism is medically unsound, or propose the germ theory of disease as an alternative. By the time we've gotten to the treatment (or the visualization design), we're so far removed from theory that there's not much we can do to correct theoretical errors. This is not to say that all of our design studies are useless for theory, or that we would make progress if we resorted purely to navel-gazing. Rather, I claim that the default shape of design study (work with the domain collaborator, do iterative design, do a post-hoc evaluation of the thing you built) won't necessarily move the needle. If we were serious about theoretical work then we should design systems with a concrete theory in mind to embody or to test. For instance, setting out at the start of a project to make an explicitly feminist~\cite{d2020data}, anarchist~\cite{keyes2019human}, or algebraically compliant~\cite{kindlmann2014algebraic} design. Or we might have to design to \textit{falsify} or \textit{attack} theories with ``reductio ad absurdum'' designs~\cite{correll2018ross} specifically meant to address common refrains of design best practices~\cite{inbar2007minimalism}. 

\section{Conclusion}
I do not intend this thought experiment to produce the conclusion that we should not evaluate our visualizations, or even that we are evaluating too much. Rather, I would charge the reader to consider if our evaluations reliably answer interesting questions, and what would need to change in the way we evaluate (or how we think of the visualization field conceptually) for this reliability to increase. What do our evaluations tell us about what we've made and how does that knowledge help us advance the field? How do we move out of the heroic age?

%\acknowledgments{
%Acknowledgments omitted for review.}

%\bibliographystyle{abbrv}
\bibliographystyle{abbrv-doi}

\bibliography{template}
\end{document}